\documentclass[a4paper,11pt]{article}
\pdfoutput=1

\usepackage{jcappub_mod}

\ifx\pdfoutput\undefined
\usepackage[dvips,bookmarks=false]{hyperref}	
\else
\usepackage{hyperref}	
\fi
\hypersetup{colorlinks,bookmarksopen,bookmarksnumbered,citecolor=blue,
linkcolor=blue,pdfstartview=FitH,urlcolor=blue}

\usepackage{amssymb} 
\usepackage{graphicx}
\usepackage{amsmath}
\usepackage{xcolor}
\usepackage{physics}
\usepackage{mathtools}
\usepackage[normalem]{ulem}

\def\beq{\begin{equation}}
\def\eeq{\end{equation}}
\def\beqa{\begin{eqnarray}}
\def\eeqa{\end{eqnarray}}

\newcommand*{\sinc}{\mathrm{sinc}}

\newcommand\JCAP[1]{\textcolor{black}{ #1}}

\title{Using Axion Miniclusters to Disentangle the Axion-photon Coupling and the Dark Matter Density}
\author[a]{Virgile Dandoy,}
\author[b]{Joerg Jaeckel,}
\author[b]{Valentina Montoya}
\affiliation[a]{Institut f\"ur Astroteilchenphysik, Karlsruhe Institute of Technology (KIT),\\ 
Hermann-von-Helmholtz-Platz 1, 76344 Eggenstein-Leopoldshafen, Germany}

\affiliation[b]{Institut f\"ur theoretische Physik, Universitat Heidelberg, Philosophenweg 16, 69120 Heidelberg, Germany}

\emailAdd{virgile.dandoy@kit.edu}
\emailAdd{jjaeckel@thphys.uni-heidelberg.de}
\emailAdd{montoya@thphys.uni-heidelberg.de}

\abstract{Dark matter direct (and indirect) detection experiments usually can only determine a specific combination of a power of the coupling and the dark matter density. This is also true for axion haloscopes which are sensitive to the product $g^{2}_{a\gamma\gamma}\rho_{\rm DM}$, the combination of axion-photon coupling squared and the dark matter density.  
In this note we show, that in the lucky case when we intersect with a so-called axion minicluster of a suitable size, we can utilize the spectral information available in haloscopes to determine the gravitational potential of the minicluster. We can then use this to measure separately the coupling and the density of the minicluster.
}

\begin{document}
\maketitle

\section{Introduction}
To establish the nature of dark matter we clearly would be happy to discover a suitable candidate. This could be found in experiments such as, e.g., searches at the LHC (cf., e.g.,~\cite{Abercrombie:2015wmb}) or at axion search experiments such as ALPS II or IAXO~\cite{Bahre:2013ywa,Spector:2016vwo,Vogel:2013bta,IAXO:2019mpb} by either producing them inside the experiment itself (e.g. LHC or ALPS II) or by looking at new particles produced in a known source (e.g. the sun in the case of IAXO). 

At the next level, of course, we would want to show that this candidate is indeed present in the Universe. Both direct and indirect detection experiments (see~\cite{Sikivie:1983ip,ADMX:2009iij,Horns:2012jf,Jaeckel:2013eha,Caldwell:2016dcw,McAllister:2017lkb,Alesini:2019ajt,Lee:2020cfj,DMRadio:2022pkf,ADMX:2023rsk,Irastorza:2018dyq,Grin:2006aw,Regis:2020fhw,Foster:2021ngm,DeRocco:2022jyq,Escudero:2023vgv, Fermi-LAT:2016afa} for some axion centered examples of both approaches) can do it (as well as at the same time discovering a suitable candidate, of course).

However, even if we have a discovery by direct or indirect detection, we are not yet sure whether the signal is caused by the dominant form of dark matter or whether it originates from a sub-dominant but more strongly coupled fraction of the dark matter.
The reason for this is that these experiments are usually only sensitive to some combination of a coupling and the dark matter density, $g^n\rho_{\rm DM}$\footnote{In direct detection the combination is usually $g^{2}\rho_{\rm DM}$, but in indirect detection it is $g^2\rho_{\rm DM}$ for decays and $g\rho_{\rm DM}$ for annihilation. However, direct and indirect detection are often sensitive to different couplings. Therefore, a combination is non-trivial even if the coupling dependencies are different.}.
We are therefore faced with a degeneracy in the measurement that makes it impossible to say how much of the dark matter we have actually discovered.
In principle, this degeneracy can be lifted if we have a signal for the same type of particle in two experiments with different dependencies on coupling and density. An example could be a particle search experiment that is independent of the dark matter density and a direct detection experiment. For WIMPs, an example of this has been investigated, e.g. in~\cite{Bertone:2010rv}, combining direct detection experiments and LHC. However, at present, for most particles there is only limited overlap in the sensitivity in couplings/masses of experiments where such a ``double detection'' is possible.

In this paper, we show an optimistic case where a measurement of the coupling might be possible with a single axion direct detection experiment, specifically a haloscope~\cite{Sikivie:1983ip,Sikivie:1985yu}.
To do so, we use the fact that these experiments can not only detect dark matter axions but they can also measure the energy spectrum with excellent precision.
It is already known that this information can be used to measure the velocity squared spectrum of the local dark matter~\cite{Foster:2017hbq,Foster:2020fln,OHare:2017yze,Irastorza:2012jq,Millar:2017eoc}\footnote{Indeed, axion dark matter experiments also have a chance to achieve directional sensitivity~\cite{Irastorza:2012jq,Jaeckel:2013sqa,Jaeckel:2015kea,Millar:2016cjp,Millar:2017eoc,OHare:2017yze,Knirck:2018knd}. }. However, if we encounter a gravitationally bound object made of axions, e.g. an axion minicluster, the resolution in the spectrum may be enough to resolve the depth of the gravitational well, such that, using the Poisson equation for the gravitational potential, we can then determine the density of the minicluster. Together with the signal strength, that is proportional to $g^{2}_{a\gamma\gamma}\rho_{\rm minicluster}$, this allows to determine the coupling $g^{2}_{a\gamma\gamma}$. After the minicluster has passed, one can then measure the signal caused by the local homogeneous density, $g^{2}_{a\gamma\gamma}\rho_{\rm DM, homogeneous}$ and therefore the corresponding density $\rho_{\rm DM, homogeneous}$. Comparisons to the expected DM density can finally be done at our location in the galaxy to check whether axions are a dominant component of the dark matter.\footnote{Such a statement can only be achieved if an ${\mathcal{O}}(1)$ fraction of the dark matter is then found to be in the homogeneous density, as suitable axion miniclusters are too rare to collect statistics on this component.}

In the following, we first start in Sec.~\ref{sec:minicluster} by briefly reviewing axion miniclusters and their properties and in Sec.~\ref{sec:haloscopesignal} the expected signals in cavity haloscopes. We explicitly demonstrate how the axion-photon coupling can be reconstructed in Sec.~\ref{sec:reconstruct} and  in Sec.~\ref{sec:rates} we quantify how lucky (admittedly very) we have to be, by discussing the rate with which we could encounter suitable miniclusters. We provide some further discussion of the results and conclusions in Sec.~\ref{sec:conclusions}.

\section{Wave Functions of Self-Gravitating Axion Miniclusters}\label{sec:minicluster}

In this section, we summarize the construction of axion miniclusters (AMC) in a wave formalism. We start by discussing in Sec.~\ref{Subsec: Random Phase Model} the general set up for the construction of the wave function of a self-gravitating system, based on a random phase assumption \cite{Widrow:1993qq} for the coefficients of each mode. In Sec.~\ref{Subsec: WKB}, the wave function of an AMC with fixed mean density and gravitational potential is introduced following Refs.~\cite{Dandoy:2022prp,Yavetz:2021pbc} (see also \cite{Lin:2018whl,Dalal:2020mjw,Widrow:1993qq}). Finally, we discuss the statistical properties of this solution.

\subsection{General Formalism and Random Phase Model}\label{Subsec: Random Phase Model}

In the non-relativistic and low density regime, axions are well described by the Schroedinger equation. More precisely, for a not too dense self gravitating object, like an AMC, we have a complex scalar field $\psi(\boldsymbol{x},t)$ obeying the Schroedinger-Poisson (SP) equations~\cite{Widrow:1993qq}, 
\begin{equation}\label{Eq:S-P system}
\begin{split}
&i\partial_t \psi = -\frac{\nabla^2}{2 m_a}\psi + m_a\phi \psi,\\
& \nabla^2 \phi = 4 \pi G m_a \lvert\psi\rvert^2 = 4 \pi G \rho ,
\end{split}
\end{equation}
where $\phi$ is the gravitational potential and $m_a$ the axion mass.
The density in the non-relativistic approximation of the axion field is $\rho = m_a \lvert\psi\rvert^2$. 

Importantly, the typical de Broglie wavelength is expected to exceed the inter-particle separation usually by a significant margin. In other words, the typical occupation numbers are very large such that $\psi$ can be viewed as a classical field describing a large number of axions ~\cite{Sikivie:1983ip,Hui:2021tkt}.

This equation has been solved extensively with numerical simulations~\cite{Li:2018kyk,Mocz:2017wlg,Schive:2014dra,Schwabe:2016rze,Schwabe:2020eac}. However, analytical approximations have been developed to reduce the computational cost while still retaining a good description of the system~\cite{Foster:2017hbq, Widrow:1993qq,Lin:2018whl,Knirck:2018knd,Dandoy:2022prp}. 
One approach~\cite{Dalal:2020mjw,Lin:2018whl,Widrow:1993qq,Dandoy:2022prp} is to decompose the wave function into energy eigenmodes of the Schroedinger equation,
\begin{equation}\label{Eq: S-equation}
\begin{split}
&\psi(\boldsymbol{x},t) = \sum_{i} a_i \psi_i(\boldsymbol{x})e^{-iE_it},\\
&\left(-\frac{\nabla^2}{2m_a}+m_a\phi(\boldsymbol{x})\right)\psi_i(\boldsymbol{x}) =E_i\psi_i(\boldsymbol{x}), 
\end{split}
\end{equation}
where $\psi_i$ are the modes with $E_i$ their corresponding energy. The coefficients $a_i$ are complex and can be found by solving the Poisson equation,
\begin{equation}\label{Eq:Poisson Equation}
\begin{split}
\nabla^2 \phi(\boldsymbol{x}) &= 4\pi G m_a \lvert\psi(\boldsymbol{x},t) \rvert^2\\
&=4\pi G m_a \sum_i \lvert a_i\rvert^2 \lvert \psi_i(\boldsymbol{x})\rvert ^2+m_a\sum_{i\neq j}a_i a_j^*\psi_i(\boldsymbol{x})\psi^*_j(\boldsymbol{x})e^{-i(E_i-E_j)t}. 
\end{split}
\end{equation}

The interference between different modes on the right hand side of Eq.~\eqref{Eq:Poisson Equation} is time dependent.
This makes it difficult to solve the Poisson equation. To overcome this issue, one usually assumes that each coefficient $a_i$, carries a different random phase \cite{Widrow:1993qq}. As a simplification, we can solve  the Poisson equation on average to obtain a fully time-independent system, 
\begin{equation}\label{Eq: S-equation}
\begin{split}
&\left(-\frac{\nabla^2}{2m_a}+m_a\phi(\boldsymbol{x})\right)\psi_i(\boldsymbol{x}) =E_i\psi_i(\boldsymbol{x}),\\
&\nabla^2 \phi(\boldsymbol{x})=  4\pi G m_a \langle\lvert\psi(\boldsymbol{x},t) \rvert^2\rangle=4\pi G m_a \sum_i \lvert a_i\rvert^2 \lvert \psi_i(\boldsymbol{x})\rvert ^2.
\end{split}
\end{equation}

\begin{figure}[t]
\centering
  \includegraphics[scale=0.35]{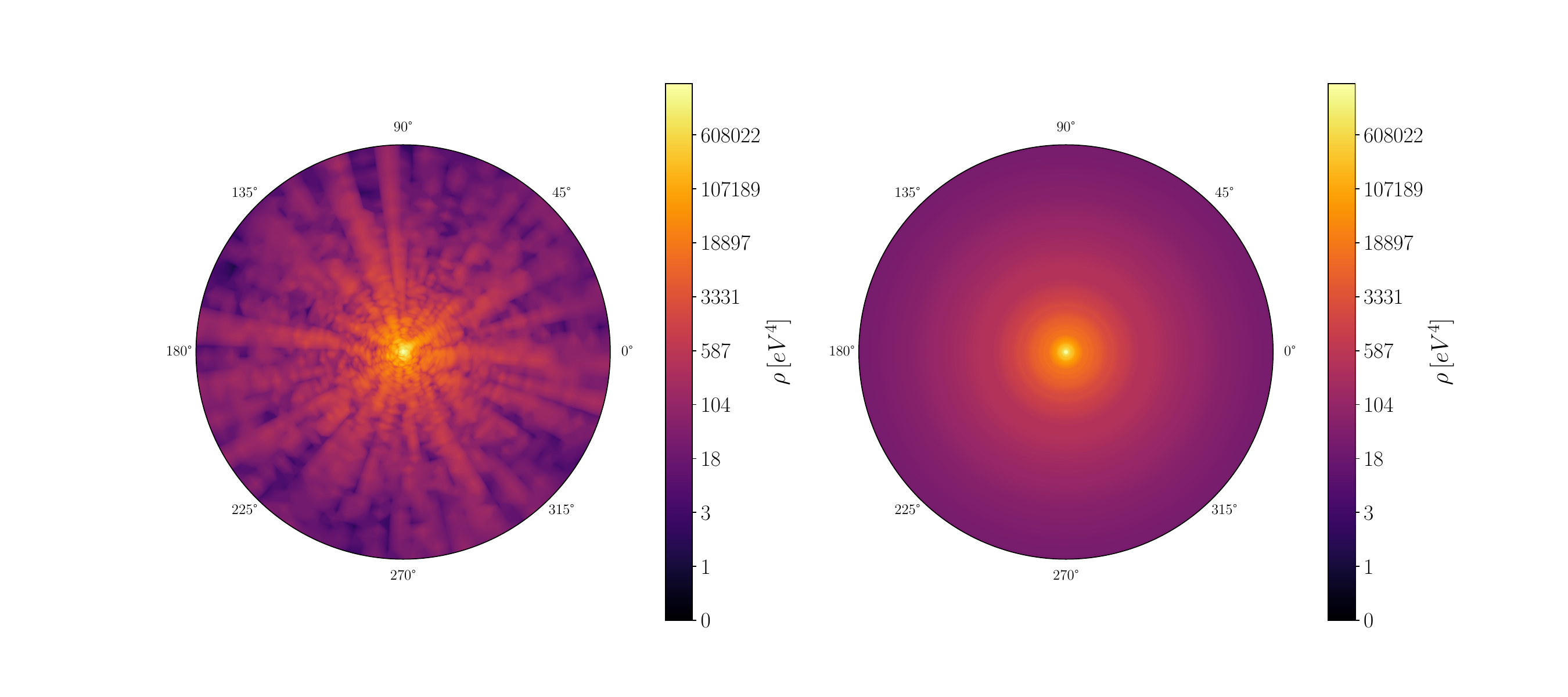}
   \caption{\textit{Left}: Single realization of the density profile of an NFW~\cite{Navarro:1995iw} AMC with mass $M=10^{-13} \, M_{\odot}$, radius $R=10^{-8}$ pc and concentration $c=10$. \textit{Right}: Density profile averaged over the random phases. Note that the radial features in the left panel are numerical artefacts. Increasing the grid as well as the number of angular momentum modes is expected to remove them but also drastically increases the computational effort required. }
   \label{Fig: Single Realization AMC}
\end{figure}

The average performed is an ensemble average. In this sense, an individual minicluster will still carry a density featuring (time-dependent) fluctuations due to the interference terms,
\begin{equation}
    \rho= m_a \lvert\psi(\boldsymbol{x},t) \rvert^2\\
= m_a \sum_i \lvert a_i\rvert^2 \lvert \psi_i(\boldsymbol{x})\rvert ^2+m_a\sum_{i\neq j}a_i a_j^*\psi_i(\boldsymbol{x})\psi^*_j(\boldsymbol{x})e^{-i(E_i-E_j)t},
\end{equation}
where the fluctuations in the AMC density profile appear in the second term of the right hand side. These ''granules'' have a characteristic length scale of the order of the de Broglie wavelength,
\begin{equation}
    \ell_{\text{gran}}\sim \lambda_{\text{dB}} \sim 1/(m_a v),
\end{equation}
and a characteristic time scale 
\begin{equation}
    T_{\text{gran.}} \sim 1/(m_a v^2),
\end{equation} 
where $v$ is the typical velocity dispersion of the cluster~\cite{Hui:2016ltb,Li:2020ryg,Yavetz:2021pbc}.
An example of a realization obtained by selecting random phases is shown in Fig.~\ref{Fig: Single Realization AMC} and clearly shows this non-uniform nature. Such features are also observed in numerical simulations~\cite{Schive:2014dra,Schive:2014hza,Woo:2008nn}.

For our purposes the granules are important, because they correspond to fluctuations in the density that limit the precision with which the coupling can be measured, cf. Sec.~\ref{sec:reconstruct}.

\subsection{Explicit Axion minicluster Wave Functions in the WKB Approximation}\label{Subsec: WKB}
In order to generate an explicit realization we still need to have the wave functions.

One approach, used in Refs.~\cite{Dandoy:2022prp, Yavetz:2021pbc}, is to
fix the average gravitational potential and hence the density profile of the AMC. Then, one can use the WKB approximation to solve Eq.~\eqref{Eq: S-equation} and to derive a general expression for the coefficients $|a_i|$. 
In the case of a spherically symmetric gravitational potential $\phi(r)$ and density profile $ \rho(r)$, the wave function becomes~\cite{Dandoy:2022prp, Yavetz:2021pbc},
\begin{equation}\label{Eq: Eigenfunctions}
\begin{split}
\psi(r,\theta,\phi) = \sum_{nlm} a_{nlm} e^{i\phi_{nlm}} R_{nl}(r)Y_{lm}(\theta,\phi),
\end{split}
\end{equation}
with $n$ standing for the principal quantum number and $l,m$, respectively, for the angular momentum and its $z$ component. The energies $E_{nl}$ of the modes are assumed to be independent of the quantum number $m$. In the continuum limit of large miniclusters, the $l$-dependence also becomes unimportant, therefore we write $E_{n}$. The radial part is defined with the WKB approximation as,
\begin{equation}\label{Eq: WKB approximaton}
\begin{split}
R_{nl}(r) = \frac{1}{\sqrt{\mathcal{N}_{nl}}}\frac{1}{r}\frac{1}{[2m_a\left(E_n-V_l(r) \right)]^{1/4}}\sin\left(\int^r dr' \sqrt{2m_a\left(E_n-V_l(r') \right)}+\pi/4\right),
\end{split}
\end{equation}
where $V_l(r)$ is the effective potential for a given angular momentum
and $\mathcal{N}_{nl}$ is a constant to assure the correct normalization. With that, the Poisson part of Eq.~\eqref{Eq: S-equation} yields~\cite{Dandoy:2022prp, Yavetz:2021pbc}, 
\begin{equation}\label{Eq:Coefficients}
 a_{nlm} = 4\pi \sqrt{m_a\mathcal{N}_{nl}\frac{f(E_n)}{g_l(E_n)}\, },    
\end{equation}
where $g_l(E_n) = 2 m_a \, \mathcal{N}_{nl}/\pi $ is the density of states for a given angular momentum $l$\footnote{Note that, we will actually never need to calculate the density of states since it will drop out when the continuous approximation for the energy levels is taken (see Ref.~\cite{Dandoy:2022prp}).}  and $f(E_n)$ is the distribution function of the AMC. The latter is directly derived from $\phi(r)$ and  $\rho(r)$ with the \JCAP{Eddington} formula~\cite{1916MNRAS..76..572E}.\\

Note that, in the classically forbidden regions, the WKB wave functions decay exponentially. For a given mode, the decay length is roughly given by its wavelength, $\lambda = 2\pi/\sqrt{2m_a(E_n-V_l)}$. To obtain a sufficient accuracy in our measurement of the gravitational potential via the energy spectrum of the axions, we have to make sure that the wave function at a given location $r$ does not receive important contribution from the exponential tail of lower (classically forbidden) energy modes with turning point at $(r-\Delta)$. 
This requires the somewhat stronger condition $\sqrt{2m_a |\phi(r)-\phi(r-\Delta)|}\Delta>>1$.
Estimating $\phi\sim m_a G M/R$ as in~\cite{Dandoy:2022prp}\footnote{This assumes that the impact parameter of our cluster crossing is not too much smaller than $R$.}, with $M$ the cluster mass and $R$ its radius, we find that this translates into $\lambda/R \ll \left( \Delta/R\right)^{3/2}$. Assuming a reasonable precision of $\Delta/R\sim 0.01$ (note that this also approximately gives the relative precision of the potential), we get the following parametric condition on the cluster parameters,
\begin{equation}\label{Eq: Parametric condition}
    1\ll 1.2\times 10^4 \left(\frac{m_a}{50\rm \mu eV}\right)\left(\frac{M}{10^{-5}\rm M_{\odot}}\right)^{1/2}\left(\frac{R}{10^{-4}\rm pc}\right)^{1/2}.
\end{equation}
\JCAP{Only clusters satisfying this inequality will be considered in this work and we indicate the corresponding constraint on the parameter space in Fig.~\ref{Fig:Rate_2D}.} 
Therefore, from now on we neglect the exponentially decaying part of the WKB wave function.

\JCAP{Let us briefly pause and summarize the physical features of this spectrum. In our approximation the axion spectrum of the minicluster features a sharp cutoff at both ends.
In the classical limit  the upper end of the spectrum can be understood by noting that all the particles with kinetic energy larger than the potential are not bound to the cluster. The lower end arises because particles that reach a certain distance from the center of the cluster cannot have energies below the value of the minicluster potential at this distance from the centre. From the quantum mechanical point of view we can consider the axions in the cluster to have a bound state spectrum featuring a cutoff at the energy at which axions escape the potential well. This gives the upper end of the spectrum. The lower end is given by the same argument as in the classical case. However, due to the quantum mechanical nature wave functions with lower energies also extend to some degree into the classically forbidden region. Therefore, at a given radius also wave functions with a lower energy than the classical cutoff will contribute. However, as we have just checked in Eq.~\eqref{Eq: Parametric condition} this effect should be small for the clusters of interest and we therefore expect a reasonably sharp end. 
This will translate into a (reasonably) sharp cut-off in the experimental signal. We will discuss this feature in more detail in the following section.
} 

\bigskip

Of course, for a given realization of the phases, the time dependent granules discussed above are present. In particular, if the phases are distributed uniformly, the constructed axion field is a Gaussian random field and the corresponding density profile, therefore, follows an exponential probability distribution~\cite{Hui:2020hbq},
\begin{equation}\label{Eq:Distribution Density}
\begin{split}
P(\rho) = \frac{1}{\rho(r)}e^{-\rho/\rho(r)},
\end{split}
\end{equation}
where $\rho(r)$ is the input density profile. The variance of the density is given by,
\begin{equation}\label{Eq:Variance Density}
\begin{split}
\sigma_{\rho}(r) =\rho(r). 
\end{split}
\end{equation}

In Fig.~\ref{Fig: Single Realization AMC}, we show the resulting density profile for both a single realization and an ensemble average of a wave function constructed according to this formalism. 
\JCAP{In the left panel, the small scale granules are clearly visible as expected from the wave interference.}

The density profile considered in this example is the \JCAP{Navarro}-Frenk-White (NFW) profile~\cite{Navarro:1995iw}, characterized by its mass $M$, radius $R$ and concentration $c$ as,
\begin{equation}\label{Eq: NFW}
    \rho(r) = \frac{M}{4\pi R^3}\frac{1}{\log(1+c)-c/(1+c)}\frac{1}{rc/R\left(1+rc/R\right)^2}.
\end{equation}
\JCAP{The corresponding distribution function $f(E)$ is given by,
\begin{equation}
    f(E) = \frac{1}{m_a^4}F_0\epsilon^{3/2}\left(1-\epsilon\right)^{-\lambda}\left(\frac{-\log(\epsilon)}{1-\epsilon}\right)^q e^p,
\end{equation}
where $\epsilon=E/m_a$ and $F_0$, $\lambda$, $p$ and $q$ are coefficients given in Tab.~2 of Ref.~\cite{Widrow:2000dm}.}

\JCAP{We note that, for spherically symmetric miniclusters in virial equilibrium\footnote{\JCAP{More precisely, this assumes "ergodicity", i.e. a system that explores the energy density in an uniform way, so that the distribution function, $f(E)$ is uniform on the energy surface of interest.}}, the distribution function depends only on the energy of the system. This results in a unique isotropic velocity distribution~\cite{2008gady.book.....B} that we will use in our analysis.   }

\JCAP{We stress that this formalism applies for miniclusters in virial equilibrium.
In the following we will consider this simplified case. We note, however, that miniclusters will be subject to stellar interactions.
The non-equilibrium state following such interactions cannot be captured by the wave function constructed in this section. 
This is particularly true for a cluster that crosses Earth itself, thereby coming close to both Earth and the Sun. A better analysis will therefore require detailed simulations of these effects, which is far beyond the scope of the present investigation. Our investigation should therefore be considered more an estimate of the relevant orders of magnitude than a precision analysis. We will comment on this again in later sections. 
}

\section{Axion miniclusters in Haloscope Experiments}\label{sec:haloscopesignal}

As a concrete case study of how to distinguish the coupling and the density by using an axion minicluster, we will focus on cavity haloscopes~\cite{Sikivie:1983ip}.
In particular, we have in mind a setup, as currently being operated by the ADMX collaboration~\cite{ADMX:2009iij,ADMX:2023rsk} (see also ORGAN \cite{McAllister:2017lkb}, QUAX \cite{Alesini:2019ajt} or CAPP \cite{Lee:2020cfj}). However, our discussion should essentially also apply to experiments such as MADMAX~\cite{Caldwell:2016dcw} or DM-Radio~\cite{DMRadio:2022pkf}, as long as they employ sufficiently spectrally resolving detection schemes (see below).

A crucial ingredient for our approach is to make use of the very high spectral resolution. As long as linear amplifier schemes (cf.~\cite{Lamoreaux:2013koa} for a discussion of such schemes vs photon counters) are used, the output signal can be Fourier transformed, allowing to analyze the signal with high spectral resolution. In practice this has been demonstrated by the ADMX collaboration~\cite{ADMX:2006kgb}.

Recently, the expected haloscope signal in the case of an axion field background with given momentum distribution and allowing for random phases has been derived in Refs.~\cite{Foster:2020fln, OHare:2017yze, Foster:2017hbq}. Similarly, we proceed here to obtain the expected signal for an AMC crossing the Earth.

As recalled in Appendices~\ref{App: Power in Haloscope } and \ref{App:AMC PSD}, following Refs.~\cite{Foster:2017hbq,Foster:2020fln}, and applied to the AMC discussed in the previous section, with field Eq.~\eqref{Eq: Eigenfunctions}, the spectral power for a finite measurement time $T$ is given by,
\begin{eqnarray}
\label{eq:powerspectrum}
    S(\omega_d)\!&\approx&  \! T \frac{\left(g_{a\gamma\gamma} B_0\right)^2 \mathcal{G}\,V}{2m_{a}}
    \\\nonumber
    &&\qquad\qquad\times   \lvert\sum_{nlm}\frac{\omega_{nlm}^2}{\left(\omega_{j}^2-\omega_{nlm}^2-i\omega_{j}\omega_{nlm}/Q\right)}
    a_{nlm}   \psi_{nlm}(\boldsymbol{x}) \, \sinc\left(\left(\omega_{nlm}-\omega_d\right)\frac{T}{2}\right)\rvert^2,
\end{eqnarray}
where $B_0$ is the magnetic field in the cavity and $V$ its volume, $\mathcal{G}$ is the usual geometry factor (see~\cite{Peng:2000hd}) of the employed cavity mode with frequency $\omega_j$ and $\omega_{nlm}$ is the energy of the modes (see App.\ref{App:AMC PSD}). Haloscope experiments are usually optimized such that it is ${\mathcal{O}}(1)$. 
Moreover, we assume here that the typical wavelength of the axion field is much larger than the size of the haloscope such that it is approximately constant in the cavity volume. The measurement is then assumed to be taken at a location $\boldsymbol{x}$ inside the cluster, in a frame with the origin at the center of the AMC.
Finally, note that the bin width of the spectral power is inversely proportional to the measurement time $T$, $\Delta \omega = 2\pi/T$, such that $\omega_d$ are \JCAP{discrete} frequencies. 

Importantly, since the axion field is a Gaussian random field (due to the random phases), the spectral power will be exponentially distributed (as already pointed out in Refs.~\cite{Foster:2017hbq,Foster:2020fln}) and follows the probability distribution,
\begin{equation}\label{Eq: Probability PSD}
\begin{split}
    P(S(\omega_d)) = \frac{1}{\Bar{S}(\omega_d)}e^{-S(\omega_d)/\Bar{S}(\omega_d)}.
\end{split}
\end{equation}
\JCAP{For a fixed AMC mean density $\rho(r)$ and gravitational potential $\phi(r)$, the  mean value $\Bar{S}(\omega_d)$ has been calculated in Appendix \ref{App:AMC PSD} and is given for high enough resolution by,\footnote{\JCAP{We would like to thank the anonymous referee for pointing out an error in the way we accounted for the relative velocity of the mini-cluster, that had significant impact on the sensitivity.}} 
\begin{equation}\label{Eq: Mean PSD}
\begin{split}
    \Bar{S}(\omega_d) &=4 \pi^2 m_a^2 \Tilde{v}_{d} \int d\theta \sin(\theta)  \,f(\Tilde{v}_{d}^2+v_c^2-2\Tilde{v}_{d}v_c\cos(\theta)) \lvert C(\Tilde{v}_{d}^2+v_c^2-2\Tilde{v}_{d}v_c\cos(\theta)) \rvert^2\,\\
    &\times \Theta\left(\sqrt{-2\phi(r)}-(\Tilde{v}_{d}^2+v_c^2-2\Tilde{v}_{d}v_c\cos(\theta))\right) \Theta\left(\Tilde{v}_{d}^2+v_c^2-2\Tilde{v}_{d}v_c\cos(\theta)\right) 
\end{split}
\end{equation}
where we recall that $f$ is the energy distribution function associated to the density profile. Moreover, $v_c$ is the velocity of the cluster relative to Earth and we define $\Tilde{v}_{d} = \sqrt{2/m_a\left(\omega_d-m_a\phi(r)-m_a\right)}$ (see also Appendix \ref{App:AMC PSD} for further details).\\
From Eq.~\eqref{Eq: Mean PSD} we can see that the signal will be limited to be in the frequency range 
\begin{equation}
    \frac{m_a}{2} v_c^2 +m_a - m_a\sqrt{-2\phi(r)}v_c\leq\omega_d \leq \frac{m_a}{2} v_c^2 +m_a + m_a\sqrt{-2\phi(r)}v_c.
\end{equation}
Hence, measuring both the starting and end points of this signal leads to a direct measurement of the potential energy $m_a\phi(r)$ and the velocity $v_c$} \footnote{Of course this measurement suffers from an error resulting from the finite bin width \JCAP{as well as from the noise/background signal}. This will be discussed in the next section. }.\\
\begin{table}[t!]
\centering
\begin{tabular}{ |p{3cm}||p{6cm}| }
 \hline
 \multicolumn{2}{|c|}{Haloscope/Axion Parameters} \\
 \hline
 \hline
 Axion &  $m_a= 50 \mu$eV \\ 
& $g_{a\gamma\gamma}=10^{-15}$GeV$^{-1}$  \\ 
 \hline
 Experiment  & $B_0=8 \rm T$  \\
 & $V = 220 \rm l$\\
 & $Q = 10^5$\\
 & $\mathcal{G} = 0.69$ ($\rm TM_{010}$ mode)\\
  & $\omega_c =m_a$\\
 \hline
 \hline
\end{tabular}
\caption{Input axion and experimental parameters used for numerical calculations in this work. We used a setup similar to ADMX, in particular, we consider a cylindrical cavity and the $\rm TM_{010}$ \cite{ADMX:2009iij} mode. We furthermore assume that the cavity is tuned in a way to have the mode frequency $\omega_j$ at the axion mass. Note however, that due to the large density of the AMC, the precise specification of the haloscope/axion parameters does not affect the coupling reconstruction and this table is only for illustration.}
\label{Tab: Parameters}
\end{table}
\begin{figure}[t]
\centering
  \includegraphics[scale=0.35]{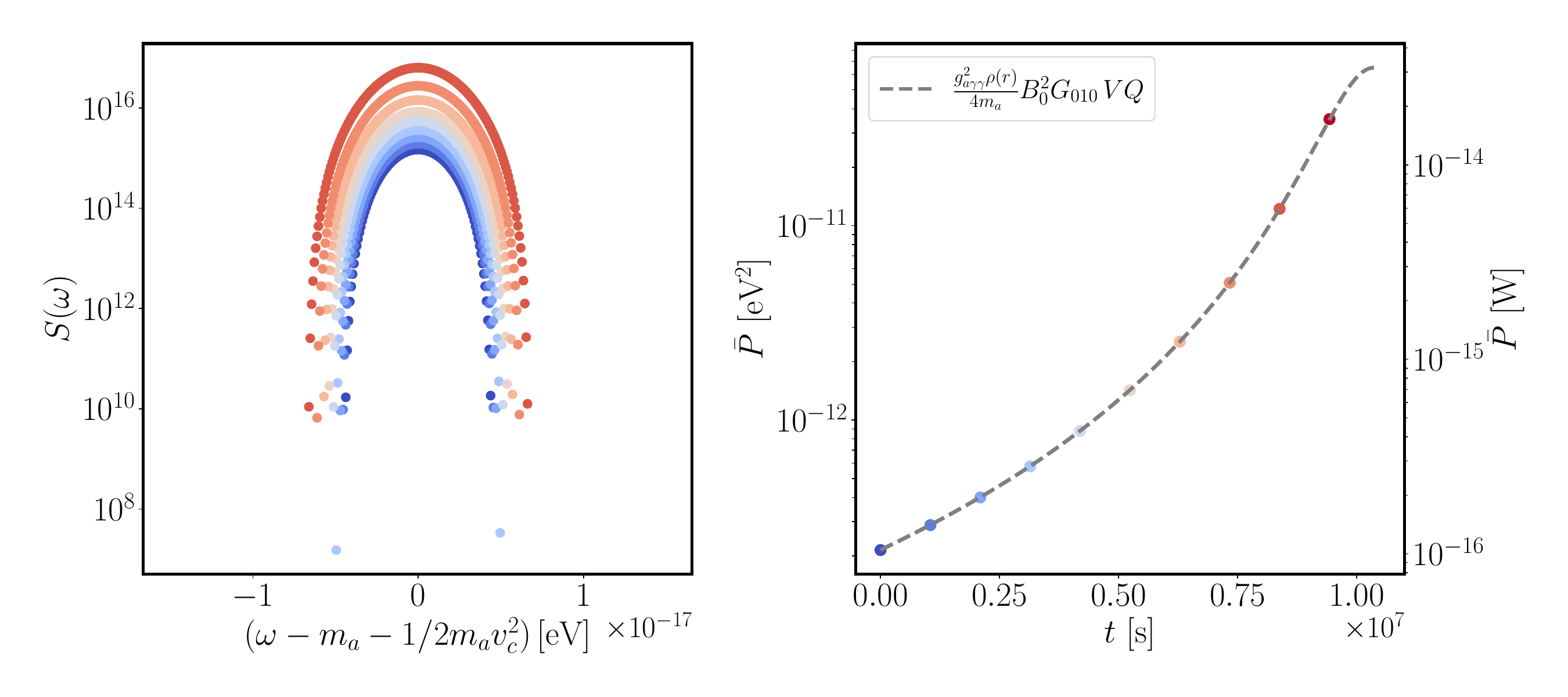}
   \caption{\JCAP{Averaged signal of an NFW profile axion minicluster crossing the Earth. For this example we consider a mass $M=10^{-10}\, M_{\odot}$, radius $R=10^{-5}$ pc and concentration $c=10$. The velocity of the AMC is $v=10^{-4}c$ and we are crossing it with an impact parameter $b=10^{-6}$ pc. We assume that each measurement period is $T=5\times10^4 \,\rm s$.  \textit{Left}: Averaged power spectral density at each measurement location (from blue to red as we are moving toward the cluster center). \JCAP{The dots are showing the power in  each frequency bin.} \textit{Right}: Averaged power as a function of time. Each dot represents the power calculated from the spectral power at a different location. The axion and cavity parameters are as in Tab.~\ref{Tab: Parameters}.}}
   \label{Fig: 1e25 Signal}
\end{figure}
From the power spectral density, the overall power induced in cavity would just be the sum over the spectral power (see Appendix \ref{App: Power in Haloscope }),
\begin{equation}\label{Eq:Power}
\begin{split}
    P &= \frac{\omega_{j}}{Q}\frac{1}{4\pi}\sum_d \Delta\omega S(\omega_d),\\
    &\approx \frac{\omega_{j}}{Q}\frac{1}{4\pi}\int d \omega S(\omega).
\end{split}
\end{equation}
Again, it is important to remember that due to the statistics of the axion field, the power is randomly distributed as well. From Eq.~\eqref{Eq: Mean PSD}, the mean power is easily calculable and is given by,
\begin{equation}\label{Eq: Mean Power}
     \Bar{P} \approx \frac{ g_{a\gamma\gamma}^2\rho(r)}{4m_a}B_0^2\mathcal{G}\,V\text{min}\left(Q,Q_a\right),
\end{equation}
such that we recover the usual result that the power is proportional to $g_{a\gamma\gamma}^2\rho(r)$ \cite{Sikivie:1983ip,Sikivie:1985yu,Berlin:2019ahk}.\\
It can be easily checked, using the central limit theorem, that the power is Gaussian distributed. Its variance, $\sigma_P$, is found to be proportional to the mean power $\Bar{P}$\footnote{Note that this is to be expected since the variance of the density $\sigma_{\rho}$ is proportional to $\rho$.} and to the inverse measurement time $\sqrt{T_{\rm gran}/T}$. Explicitly,
\begin{equation}\label{Eq: Variance Power}
\begin{split}
    \frac{\sigma_P}{\Bar{P}}&\sim \sqrt{\frac{2\pi}{T}\frac{1}{ m_a\phi(r)}},\\
    &\sim \sqrt{\frac{T_{\text{gran}}}{T}}.
\end{split}
\end{equation}
We recall that the time scale of the granules was defined as $T_{\text{gran}}\sim 1/(m_a v^2)\sim 1/(m_{a}\phi)$, where $v$ is the velocity dispersion in the cluster at radius $r$.

We show in Fig.~\ref{Fig: 1e25 Signal} an illustrative example of mean spectral power (left) and integrated power (right) for axion and haloscope parameters described in Tab.~\ref{Tab: Parameters}. \JCAP{This signal is for an NFW cluster (see Eq.~\eqref{Eq: NFW}) with mass, radius and concentration respectively given by $(10^{-10} M_{\odot} , 10^{-5} \, \text{pc}, 10 )$. The Earth is assumed to cross the minicluster with a relative velocity of $10^{-4}c$ and an impact parameter $b=10^{-6}$ pc. The relatively low value of the relative velocity was originally motivated by the necessity of having enough time to measure at multiple locations along the path throughout the minicluster. However, it turns out that also higher velocities usually work well, since the observable width of the spectrum also increases with this velocity. That said, in Sec.~\ref{sec:rates}, we will consider a realistic distribution for the relative velocity of the miniclusters thereby including also higher relative velocities. In the left panel, each colored line represents the mean spectral power at a different location inside the cluster (from blue to red). At each location, the measurement time is taken to be $T=5\times 10^4\,  \rm s$, leading to a bin width $\Delta \omega \approx 10^{-19} \rm eV$. In the right panel, the corresponding mean power has been calculated at each location (colored dots). We furthermore carefully checked, that the approximation given in Eq.~\eqref{Eq: Mean Power} was in line with the exact expression for the power in Eq.~\eqref{Eq:Power}. 
Finally, since the cluster we have considered has a density more than $\sim 6$ orders of magnitude larger than the expected dark matter density background, it is not surprising to find that the power gets the same scale difference compared to the background axion field power $P_{\rm background}\sim 10^{-22}\,  \rm W$ \cite{OHare:2017yze,Sikivie:1983ip,Sikivie:1985yu,ADMX:2023rsk}.} 
\section{Reconstruction of the Axion-Photon Coupling}\label{sec:reconstruct}
As emphasized, the spectral power offers the promise of a direct measurement of the axion minicluster gravitational potential. This is the key for the axion-photon coupling reconstruction, since from it, and with the use of the Poisson equation, we have access to the density of the cluster. The coupling and the density are then no longer entangled in the power measurements.\\
In this section, we proceed to construct a formalism based on the Poisson equation, that enables us to disentangle the coupling from the density as just described. We then apply this method on simulated signals in order to delineate the region in the AMC parameter space where a reconstruction can be successful. 

\subsection{General Method}

Let us first of all declare the assumptions used in the following calculations. We consider a similar hypothetical scenario as considered in Ref.~\cite{OHare:2017yze} where the axion and its mass have been found after some scan over a wider mass range. Once this is done, the axion mass is no longer an unknown parameter. Secondly, following the discussion initiated around Fig.~\ref{Fig: 1e25 Signal}, the typical signal induced by an AMC is orders of magnitude higher than in the case of a background axion field signal. We therefore expect that the thermal and quantum noises~\cite{Berlin:2019ahk, Clerk:2008tlb} are negligible in our case.
\JCAP{Moreover, we also use a simplified setup where the minicluster is fully virialized and is unperturbed by the gravitational potential of the Earth and the Sun and Earth crosses the minicluster on a straight line. This is clearly not realistic. However, we nevertheless hope that this treatment captures the relevant order of magnitudes for the measurement and that a more realistic situation can be addressed with numerical simulation of the cluster encounter (cf, e.g.~\cite{Witte:2022cjj}, for an example of a minicluster in the vicinity of a neutron star) and a corresponding more careful analysis of the time dependence of the signal.}

As pointed out in the previous section, the \JCAP{width of the spectral power $S(\omega)$} provides a direct measurement of the gravitational energy $m_a \phi(r)$. Since the AMC is a self-gravitating object, its density $\rho(r)$ and gravitational potential $\phi(r)$ are directly related via the Poisson equation,
\begin{equation}
    \nabla^2 \phi(r) = 4 \pi G \rho(r).
\end{equation}
However, note that what is actually obtained from the spectral power is the gravitational potential as {\emph{function of the measurement time}} $t$. In order to make use of the Poisson equation, we then have to parameterize the radial motion of the Earth throughout the AMC. This can be done via \footnote{\JCAP{Note, again, that a realistic description of the Earth' motion should account for its orbit around the sun. In this more realistic picture, the trajectory would therefore no longer be a straight line. However, the velocity of the Earth around the sun ($\sim 30\,  \rm km/s $) is  smaller than the typical velocity of the minicluster ($\sim 300\,  \rm km/s $), therefore we consider the straight line limit as a reasonable approximation.}}, 
\begin{equation}\label{Eq: Radial motion inside AMC}
    r(t) = \sqrt{b^2 + \left(vt - \sqrt{R^2-b^2}\right)^2},
\end{equation}
where $v$ is the velocity of the Earth (in a frame with the origin at the center of the AMC), $b$ is the impact parameter and $R$ the radius\footnote{We assume here that $R=r(0)$.}. With that, the Poisson equation can be transformed by using the time $t$ as the variable, 
\begin{equation}\label{Eq: Time Poisson equation}
    \frac{\Ddot{\phi}(t)}{\Dot{r}(t)^2}+\frac{2\Dot{\phi}(t)}{\Dot{r}(t)r(t)}-\frac{\Ddot{r(t)}\Dot{\phi}(t)}{\Dot{r}(t)^3} = 4\pi G \rho(t).
\end{equation}
The procedure to reconstruct the axion-photon coupling is the following:
\begin{itemize}
    \item $\phi_{\text{out}}(t_i)$ and $(g_{a\gamma\gamma}^2\rho(t_i))_{\text{out}}$ are extracted from the power spectral density and power, respectively, at $N$ different measurement times $i$ (corresponding to $N$ different locations in the cluster). Moreover, the velocity $v_c$ of the AMC can be determined from the power spectral density as well.
    
\item We then use the Poisson equation in the form of of Eq.~\eqref{Eq: Time Poisson equation} to define the function
\begin{equation}\label{Eq: Likelihood_1}
    \mathcal{F}(b,R,g_{a\gamma\gamma}; t_i) = \frac{g_{a\gamma\gamma}^2}{4\pi G}\left(\frac{\Ddot{\phi}(t_i)}{\Dot{r}( t_i)^2}+\frac{2\Dot{\phi}(t_i)}{\Dot{r}( t_i)r( t_i)}-\frac{\Ddot{r( t_i)}\Dot{\phi}(t_i)}{\Dot{r}( t_i)^3}\right),
\end{equation}
which returns $g_{a\gamma\gamma}^2\rho(r)$ if the impact parameter, radius and gravitational potential are perfectly known.\\

\item Finally, the parameters $b$, $R$ and $g_{a\gamma\gamma}$ are reconstructed by maximizing the function,
\begin{equation}\label{Eq: Likelihood_2}
    \mathcal{L}(b,R,g_{a\gamma\gamma}) = \sum_i \log(\frac{(g_{a\gamma\gamma}^2\rho(t_i))_{\text{out}}}{\lvert(g_{a\gamma\gamma}^2\rho(t_i))_{\text{out}} - \mathcal{F}(b,R,g_{a\gamma\gamma}; t_i) \rvert}).
\end{equation}
\JCAP{Note that the choice of the maximizing function is not unique and different choices may alter the reconstruction of the parameters.}
\end{itemize}

\noindent 
Before moving to concrete applications of this formalism, let us clarify the different sources of errors associated to it.

First, the reconstruction suffers from the inevitable statistical fluctuations of the axion field (see Sec.~\ref{sec:minicluster}). Indeed, we have already mentioned that the random phases in the AMC wave function generate granule fluctuations for a specific realization of the phases. It is important to stress that, those granules do not affect the width of the power spectral density but instead translate into fluctuations in the power as discussed in Eq.~\eqref{Eq: Variance Power}. Hence, using the formalism constructed in this section, the granule structures affect the quantity $(g^2_{a\gamma\gamma}\rho)_{\rm out}$ extracted from the power measurement and those statistical fluctuations will also impact on the reconstruction of the axion-photon coupling via the maximization of Eq.~\eqref{Eq: Likelihood_2}.\par
Although the width of the power spectral density does not suffer from the axion field statistics, it naturally gets errors from the finite frequency binning $\Delta \omega = 2\pi/T$. \JCAP{The determination of the width of the signal, leading to the reconstruction of the potential $m_a\phi_{\text{out}}$, is therefore known with an uncertainty proportional to $\Delta \omega$}. This error propagates on the first and second derivatives of the potential needed to calculated the function $\mathcal{F}(b,R,g_{a\gamma\gamma})$. Thus, the reconstruction of the axion-photon coupling applying Eq.~\eqref{Eq: Likelihood_2} suffers from a systematic error which gets stronger as \JCAP{the ratio of $\Delta \omega$ over the signal width increases.  In other words, we require $\Delta \omega /(2 m_a v_c \sqrt{-2 \phi(r)}) \ll 1$. In addition to the finite binning error, the noise/background is affecting the measure of the signal width and therefore the gravitational potential reconstruction. In what follows, only the values of $S(\omega_d)$ exceeding the noise/background are considered. Although we find that this does not have a dramatic effect on the reconstruction of the potential, it will eventually degrade the measurement of the signal width as we go to lighter clusters.}

\par
It is furthermore expected, that the number $N$ of time data points $t_i$ provides 
another source of systematic error. Indeed in the limit of small $N$, the potential is only reconstructed at few locations and, similarly as before, its first and second derivatives gets less accurately reconstructed, leading to the same conclusion as before.

\subsection{Application \JCAP{to} Simulated Data}\label{Subsec: Simulations}
\begin{figure}[t]
\centering
  \includegraphics[scale=0.35]{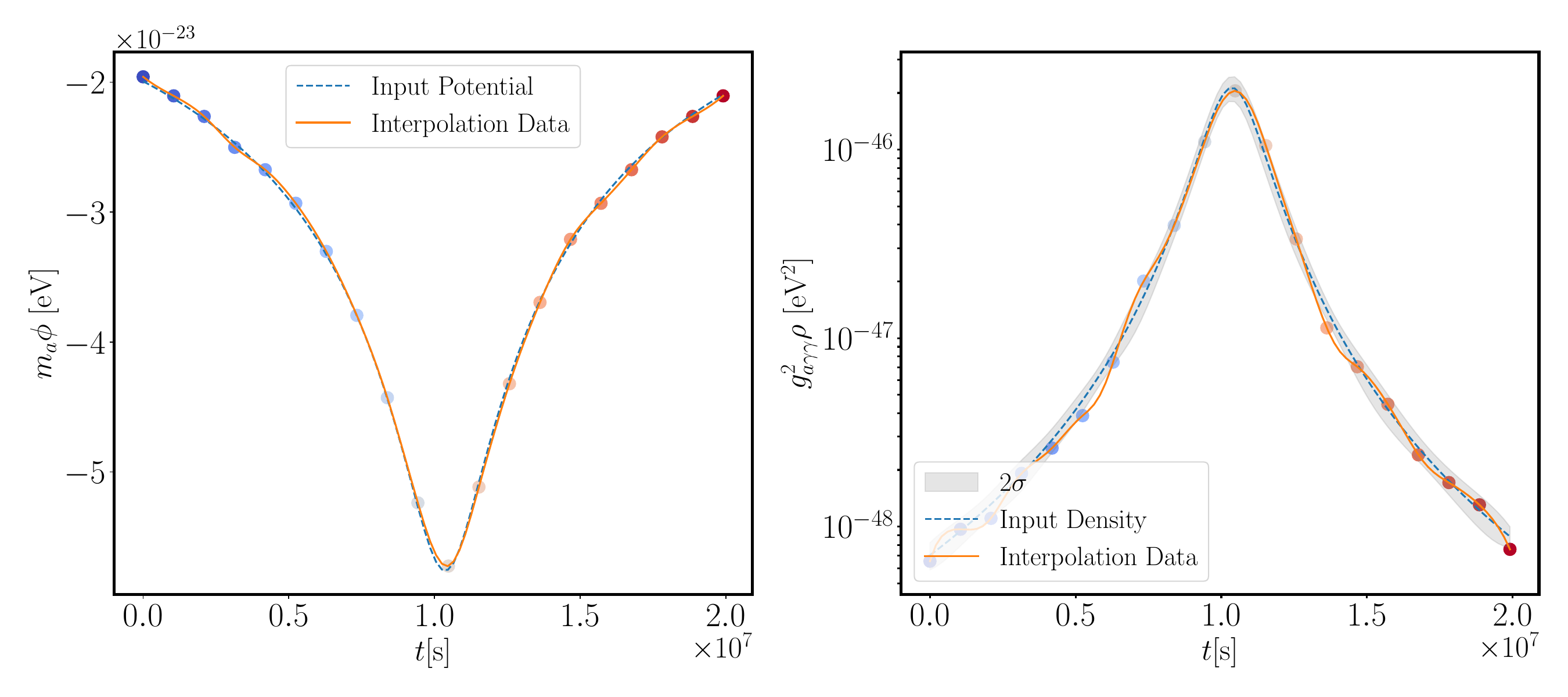}
\caption{ Reconstruction of the gravitational energy (left) and $g_{a\gamma\gamma}^2\rho(r)$ (right) for a simulated signal from a cluster characterized by \JCAP{ $M=10^{-10}\, \rm M_{\odot}$, $R=10^{-5}\, \rm pc$ and $b=10^{-6}\, \rm pc$. The number of time data points taken during the AMC crossing is $N=20$ }. In the right panel, the shaded gray region shows the expected variance $\sigma_P$ of the power (see Eq.~\eqref{Eq: Variance Power}) due to the granules of the AMC. }
\label{Fig:SimulationsI}
\end{figure}
We now proceed to apply our reconstruction formalism to concrete simulations. For our practical computations we consider encounters between an AMC featuring an NFW profile (see Eq.~\eqref{Eq: NFW}) and a detector on Earth. From the point of view of the detector the density profile, the impact parameter and the velocity of the AMC are, of course, assumed to be unknown parameters.\\
The spectral power, $S(\omega)$, induced by the AMC is simulated according to the probability distribution in Eq.~\eqref{Eq: Probability PSD} at each measurement location inside the cluster, corresponding to measurement times $t_i$. The period of the measurement $T$, defines the bin width of the spectral power. For each simulated spectral power, the induced power is finally calculated according to Eq.~\eqref{Eq:Power}. The data are at the end composed of $N$ successive measurements of $\{S(\omega;t_i),P(t_i)\}$ for $i=0,...N$.\\

 \JCAP{In Fig.~\ref{Fig:SimulationsI}, we show for a simulated cluster with mass $M=10^{-10} \, \rm M_{\odot}$, radius $R=10^{-5}$~pc and concentration $c=10$, the reconstruction of the gravitational energy $\left(m_a\phi(t_i)\right)_{\rm out}$ and $\left(g_{a\gamma\gamma}^2\rho(t_i)\right)_{\rm out}$.}
 The impact parameter and the cluster velocity are, as before, $b=10^{-6}$~pc and $v=10^{-4}c$. At each location the measurements are taken during $T=5\times 10^{4}$s and we collect data at a total of \JCAP{$N=20$ locations}. The input axion mass and coupling, as well as the experiment parameters, are taken from Tab.~\ref{Tab: Parameters}.\\
\begin{figure}[t]
\centering
  \includegraphics[scale=0.35]{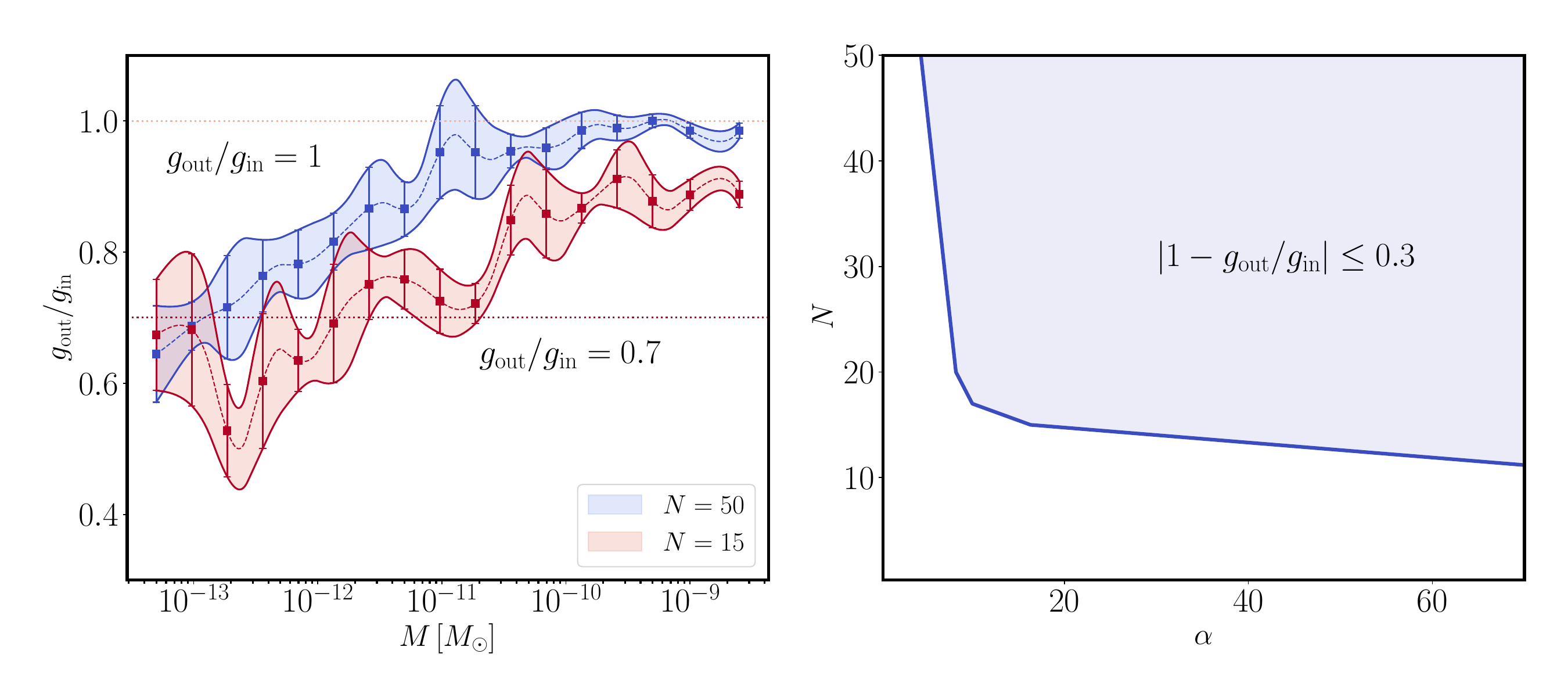}
\caption{\textit{Left}: Reconstruction of the axion-photon coupling as a function of the mass for $N=50$ and $N=15$ data points. The shaded region shows the dispersion of the reconstruction over 10 simulated reconstructions, the dashed lines the average. For both curves, the radius is set to $R=10^{-5}$ pc, the AMC velocity to $v=10^{-4}$c and the measurement time to $T=5\times 10^4$s. \textit{Right}: Sensitivity of the coupling reconstruction as a function of the number of time data points $N\approx 2R/(v T)\sqrt{1-(b/R)^2}$ and the averaged ratio of the bin and signal widths, \JCAP{$\alpha = m_{a} v_c \sqrt{2GM\frac{\log{(R/b)}}{(R-b)}}T/\pi $}. The blue shaded region shows the region where the axion-photon coupling is reconstructed with an error less than $30\%$. The dashed orange lines show the rectangle approximations used to infer the rate of encounters in Sec.~\ref{sec:rates}.}
\label{Fig:g_out}
\end{figure}

In order to illustrate the sensitivity of the construction on the minicluster mass, we show in the left panel of Fig.~\ref{Fig:g_out} the ratio between the reconstructed coupling, $g_{\rm out}$, and the input one, $g_{\rm in}$, as a function of the AMC mass. The other parameters $(v,R,b,T)$ are respectively fixed to be \JCAP{$\left(10^{-4}c,10^{-5} \text{pc}, 10^{-6} \text{pc},4\times 10^5 \text{s}\right)$.} For each mass, $10$ realizations of the signal have been simulated and the coupling has been reconstructed for each of them. The dashed lines show the average reconstructed coupling, and the solid upper and lower lines the variance of the reconstruction. Following the expectations, as the mass decreases, the ratio \JCAP{$\Delta \omega /(2 m_a v_c \sqrt{-2 \phi(r)})$} becomes larger and the reconstruction of the gravitational potential suffers from larger deviations due to the bin width $\Delta \omega$. The reconstruction procedure via the Poisson equation is then less efficient, and \JCAP{there is an increasing systematic error visible in the left panel of  Fig.~\ref{Fig:g_out} as an overall deviation from $g_{\rm out} / g_{\rm in}=1$ towards smaller values}. \JCAP{We can understand this from the fact that the reconstructed gravitational potential suffers from larger fluctuations. As the density cannot be negative the Poisson equation will then tend to overestimate the density. In order to match to the reconstructed $(g_{a\gamma\gamma}^2\rho(t_i))_{\text{out}}$, the coupling is therefore systematically underestimated.}
\JCAP{In addition to this, the power suffers from sizeable statistical fluctuations. This effect can be seen as an increase of the variance as we go to lower masses. Finally, the red and blue curves show how the number of location data points alter the reconstruction. In particular, the red and blue lines have been simulated for $N=15$ and $N=50$ points, respectively}. \par
\JCAP{Since the main influence on the reconstruction is coming from the ratio} \linebreak \JCAP{$\Delta \omega /(2 m_a v_c \sqrt{-2 \phi(r)})$ and the number of location data points $N$, we show in the right panel of Fig.~\ref{Fig:g_out}, the region of those parameters that returns a reasonable reconstruction with precision $\lvert 1-g_{\rm out} / g_{\rm in} \rvert \leq 0.3$. In particular, the number of data points has been expressed as }\JCAP{the ratio between the total crossing distance and the distance per measurement,  $N=2R\sqrt{1-(b/R)^2}/(v T)$, assuming that each point is taken after having measured at a location during a period $T$. Moreover, since the quantity \JCAP{$\Delta \omega /(2 m_a v_c \sqrt{-2 \phi(r)})$} depends on the location where the measurement is performed, we use in Fig.~\ref{Fig:g_out} the potential averaged over the path of the Earth throughout the AMC. The resolution of the gravitational potential will therefore be approximated as $\alpha \equiv m_{a} v_c \sqrt{2GM\frac{\log{(R/b)}}{(R-b)}}T/\pi $. }

\section{Rate of Encountering Suitable AMCs}\label{sec:rates}

In this section, we estimate the rate at which we may encounter an AMC with parameters that allow for a reasonable reconstruction of the axion-photon coupling (see blue region in right panel of Fig.~\ref{Fig:g_out}). Our approach is somewhat simplistic. We assume that all axion miniclusters are spherically symmetric, with  the same size and have the same mass. \footnote{\JCAP{A better estimate for the rate should take into account a more realistic mass distribution for the AMCs. See for instance Ref.~\cite{Pierobon:2023ozb}. In light of our relatively crude approximations made for several other effects we neglect this effect.}} The rate is then given by, 
\begin{align}
    \Gamma = n_{\rm AMC}(r) \langle \sigma v \rangle,
\end{align}
where,
\begin{align}
    n_{\rm AMC}(r) = f_{\rm AMC}\frac{\rho_{\rm DM}(r)}{ M},
\end{align}
and $f_{\rm AMC}$ is the fraction of the total dark matter density in AMCs \JCAP{and $M$ is the AMC mass.}

The local density of dark matter is modeled by an NFW profile evaluated at $r = 8.33$ kpc, 
\begin{align}
     \rho_{ \rm DM}(r) = \frac{\rho_{s}}{(r/r_{s})(1 + r/r_{s})^{2}},
\end{align}
with $\rho_{s}= 0.014 \rm M_{\odot} pc^{-3}$, $r_{s}= 16.1$~kpc \cite{Nesti:2013uwa}, so that the local dark matter density is 
\begin{equation}
\rho_{\rm DM} = 0.45\,  \rm GeV/cm^3.
\end{equation}
 
Concerning the minicluster fraction, $f_{\rm AMC}$,  Refs.~\cite{Eggemeier:2019khm,Ellis:2020gtq, Pierobon:2023ozb} for example, find from numerical simulations that the fraction of axions bound in AMCs is $\sim 0.75$ at redshift around $z\sim 100$. However, it is quite uncertain how this evolves until today. 
In any case, the numerical values in our figures show rates divided by $f_{\rm AMC}$. But to give rough and optimistic numbers we assume $f_{\rm AMC}\sim 1$. 

The geometrical cross section to encounter an AMC with impact parameter less than $b$ \JCAP{and a relative velocity between $v_{i}$ and $v_{f}$} is given by $ \sigma(b) = \pi b^{2}$ and the differential rate becomes,
\begin{align}\label{Eq: diff Rate}
    \frac{d\Gamma}{db} &= n_{\rm AMC}(r) \left\langle  v \frac{d\sigma}{db}  \right\rangle, \nonumber\\
 &= n_{\rm AMC}(r) \int_{v_i}^{v_f} v f(v) \frac{d\sigma}{db} dv,
\end{align}
where $f(v)$ is the velocity distribution. For the latter we consider the distribution in the laboratory frame from~\cite{Foster:2020fln} adapting  the conventional Standard Halo Model (SHM)  that yields to a Maxwell-Boltzmann distribution in the galactic frame, 
\begin{align}\label{Eq: Velicity distribution}
  f_{\rm gal}(v)= 4 \pi \frac{1}{\pi \sqrt{\pi}}\frac{v^{2}}{v^{3}_{0}}e^{-v^{2}/v_{0}^{2}},
\end{align}
where $v_0\sim 220\, \rm km/s$~\footnote{Some papers give slightly different velocities, e.g. Ref.~\cite{Choi:2013eda} has a value of $\sim 270 \rm km/s$. We have checked that this does not drastically alter the rate.} is the velocity dispersion~\cite{Herzog-Arbeitman:2017fte,Freese:2012xd}.
Note that, the velocity distribution is expected to be additionally cut-off beyond the escape velocity $v_{\rm esc} \sim 544 \, \rm km/s$ \cite{Blennow:2015oea}.

\JCAP{Performing the transformation to the detector rest frame~\cite{Blennow:2015oea} as $\mathbf{v} \rightarrow \mathbf{v}  - \mathbf{v}_{\rm lab}(t)$, and averaging over the spatial angles yields to the detector frame speed distribution~\cite{phdthesis},} 

\begin{align}
    \begin{dcases}
          \frac{2 v}{\sqrt{\pi}v_{0}v_{\rm lab}} e^{-v_{\rm lab}^{2}/v_{0}^{2}}\sinh\left( \frac{2v_{\rm lab}}{v_{0}^{2}} v\right) e^{-v^{2}/ v_{0}^{2}} & |\mathbf{v} + \mathbf{v_s}| < v_{esc} \\
        0 &  |\mathbf{v} + \mathbf{v_s}| > v_{esc}  \\
    \end{dcases}
\end{align}
where $v_{\rm lab}\sim 235\, \rm km/s$ is the laboratory velocity relative to the galactic frame   \cite{Sch_nrich_2010} and $v_s$ is the Sun's velocity relative to the galactic frame.
We can indeed use an angular averaged distribution, as does not matter from which direction we encounter our (spherically symmetric) AMCs.
\begin{center}
    \begin{figure}[t!]
        \centering
        \includegraphics[scale=0.35]{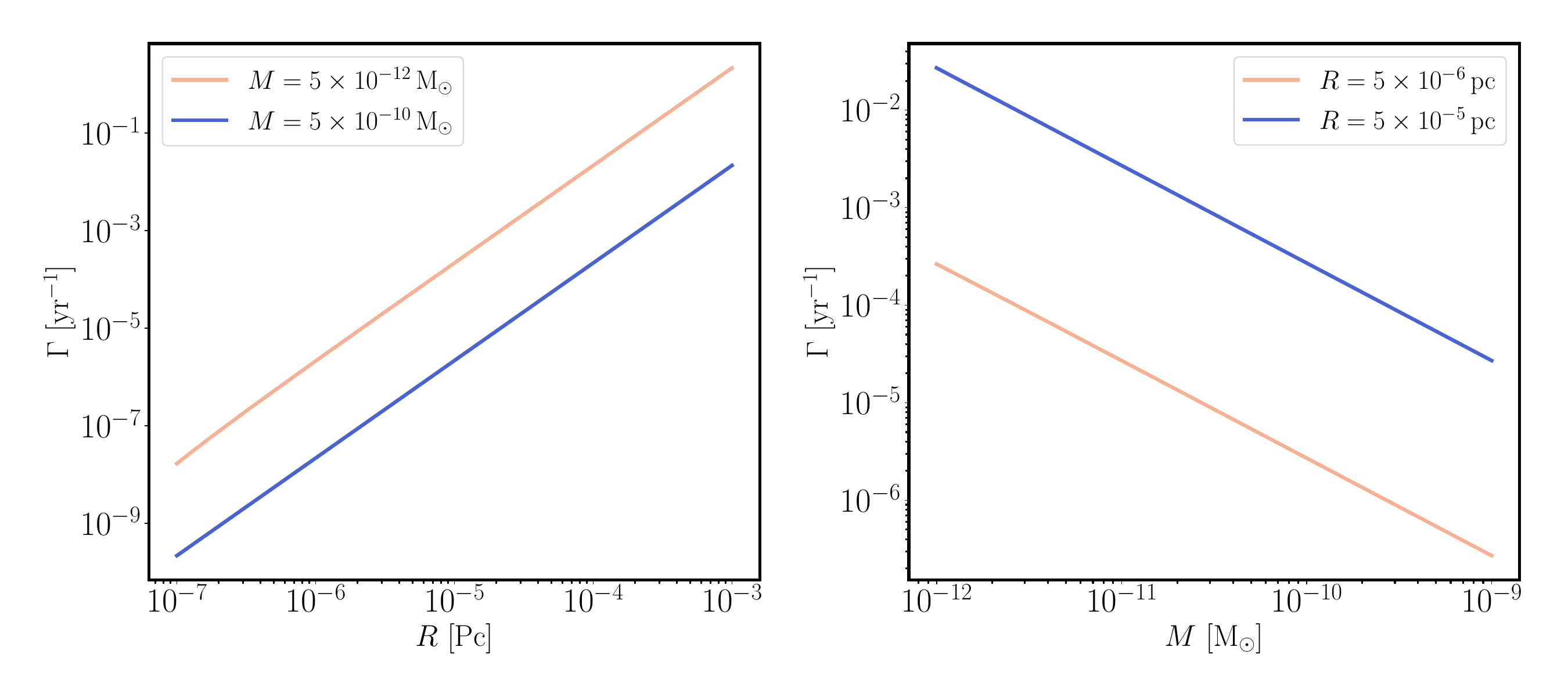}
        \caption{\textit{Left}: \JCAP{Detection rate as a function of the radius for two AMC masses. \textit{Right}: Detection rate as a function of the mass for two AMC radius. In both panels, the rate is normalized with the axion fraction bound in AMCs, $f_{\rm AMC}$.}}
        \label{fig:rates_rm}
    \end{figure}

\end{center}
\JCAP{According to Fig.~\ref{Fig:g_out}, we estimate that a reasonable axion-photon coupling reconstruction is possible if 
\begin{align}
     \alpha_{\rm min} &\leq m_a v \frac{T}{\pi}\sqrt{2 G   M \frac{\log{(R/b)}}{(R-b)}},\\
 N_{\rm min} &\leq \frac{2R}{vT}\sqrt{1-(b/R)^2}
\end{align}
with $\alpha_{\rm min}\approx 19$ and $N_{\rm min}\approx 17$ (these values are obtained from approximating the blue shaded region in Fig.~\ref{Fig:g_out} as a rectangle).}  \JCAP{These constraints come from imposing a sufficiently good resolution of the gravitational potential and on enough of data points, as given by the blue region in the right panel of Fig.~\ref{Fig:g_out}}\footnote{Note, again that his "rectangular" approximation provides an easy-to-handle approximation of the blue shaded region in the left panel of Fig.~\ref{Fig:g_out}. This makes it simple to obtain a first estimate for the rate. Considering the exact shape of this surface is expected to not alter the results drastically and would lead an increase of the encounter rate}. We furthermore define $\kappa \equiv b/R$, and the maximum measurement time as $T_{\rm max}(R,b,v) = 2\sqrt{R^{2}- b^{2}}/ (vN_{\rm min})$, such that the two previous equations now become 

\JCAP{\begin{align}
     \alpha_{\rm min}N_{\rm min} \leq 2m_a/\pi \sqrt{2MGR} \sqrt{-(1+\kappa) \log({\kappa})}
    \label{ec:kappa}
\end{align}}
Solving this last equation numerically gives  the maximal impact parameter - radius ratio, \JCAP{ $\kappa_{\rm max} (M,R)$, as a function of the AMC mass and radius}. \JCAP{Note that the reconstructable impact parameters do not depend on the relative velocity. We can therefore use the full available range of velocities from $0$ to $v_{ esc}$.} Finally, using Eq.~\eqref{Eq: diff Rate} we obtain that the rate of AMC encounters that allows for a reasonable reconstruction of the axion-photon coupling is given by,
\JCAP{\begin{equation}
     \Gamma(M, R) = n_{\rm AMC}(r) \pi R^{2} \kappa_{\rm max}(M,R)^{2}\int_{0}^{v_{esc}}  v f(v)  dv.
\end{equation}}

We show in Fig.~\ref{fig:rates_rm} the resulting rate as a function of the radius (left panel) and the mass (right panel).  \JCAP{We observe that the rate typically increases as we go to larger radius. On the other hand, increasing 
the mass decreases the rate since in that case the number density of miniclusters decreases.} \JCAP{Taking the full shape of the  blue shaded region in Fig.~\ref{Fig:g_out} into account (and not the rectangle approximation considered above) would increase the rate compared to what we are presenting here.}

\begin{figure}[t]
\centering
  \includegraphics[scale=0.38]{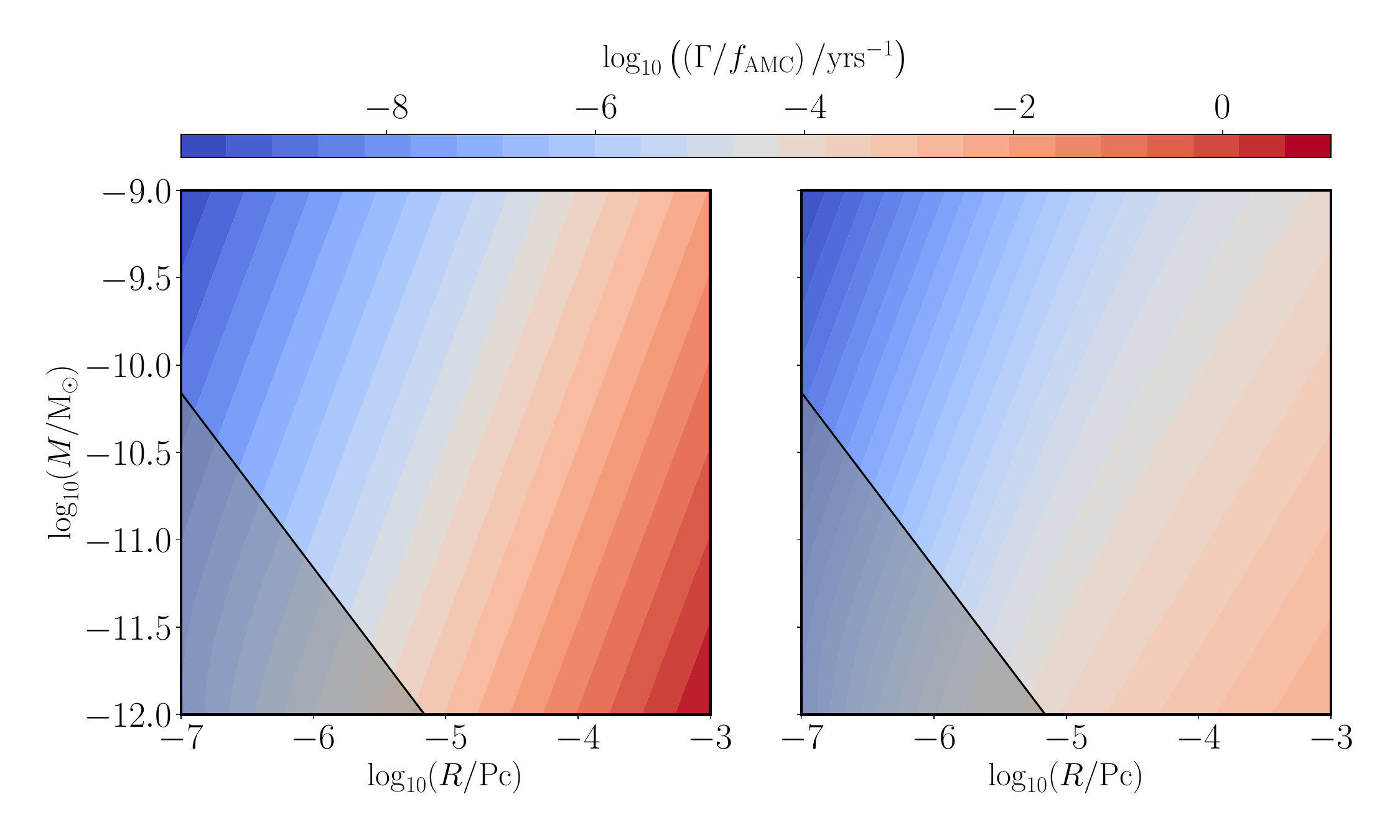}
\caption{\textit{Left}: Rate of AMC encounters (normalized by the DM fraction in AMC $f_{\rm AMC}$) that allow for a reconstruction of the axion-photon coupling. \textit{Right}: Rate accounting for the survival probability decreasing with the mean density $\sim M/R^3$ (given in Ref.~\cite{Dandoy:2022prp}). The gray shaded region corresponds to the parameter space where ou formalism does not hold, see Eq.\eqref{Eq: Parametric condition}.}
\label{Fig:Rate_2D}
\end{figure}
In Fig.~\ref{Fig:Rate_2D}, we show the rate of the AMC encounters for which a reasonable reconstruction is possible as a function of both the mass and the radius (left panel).  
\JCAP{As the mass decreases the reconstruction becomes less efficient. However, the number density of miniclusters increases even faster and the overall rate increases.} We also observe that larger AMCs have the best chances to be detected and to lead to a good axion-photon coupling reconstruction. Nonetheless, as shown in recent works (see Refs.~\cite{Kavanagh:2020gcy,Dandoy:2022prp,Shen:2022ltx,Tinyakov:2015cgg,Dokuchaev:2017psd}), AMCs might be strongly affected by tidal interactions with galactic stars.\footnote{\JCAP{Here, we are concerned with potential destructive encounters with stars other than the sun before the AMC is in our vicinity. }} Although the survival of the AMCs depends on their density profile shape (NFW, power law, etc.), it has been argued that their survival is directly proportional to the mean density of the cluster. For this reason, even if encounters with larger AMCs are expected to be more frequent, their survival should be more strongly affected compared to smaller AMCs. We show on the right panel of Fig.~\ref{Fig:Rate_2D}
the AMC encounter rate weighted by the survival probability. Note, however, that the survival probability has been extracted from the results of Ref.~\cite{Dandoy:2022prp} for a Lane-Emden profile (see for instance Ref.~\cite{Galactic}). However, it has also been argued in Ref.~\cite{Dandoy:2022prp} that clusters with density profiles similar to NFW should be more resistant to stellar interactions.

\JCAP{It is also important to consider the disruption effects that can be produced by the Earth and by the Sun. To get an estimate, lets consider the energy injection generated by an interaction with an astrophysical object, \cite{Kavanagh:2020gcy} }
\JCAP{
\begin{equation}
    \Delta E \approx \left(\frac{2GM_d}{b^2 V_d} \right)^2 \frac{M \langle R^2 \rangle}{3},
\end{equation}
where $M_d$ and $V_d$ are the mass and the relative velocity of the disturber interacting with the minicluster with mass $M$. $b$ is the impact parameter and $\langle R^2 \rangle $ is the mean squared radius of the minicluster.}

\JCAP{In practice, the amount of energy needed to destroy the minicluster is around $\Delta E \sim E_{b}$, with $E_b$ the binding energy of the system. For the disruption caused by the sun, we find that all of the parameter space shown in Fig.~\ref{Fig:Rate_2D} will suffer from tidal disruption due to the sun. 
This effect should be smaller only at higher masses and relatively small radii. As an example, for a minicluster of radius $R \sim 3\times 10^{-6}$~pc the minimum mass to survive the Sun disruption is $M \sim 10^{-4} \text{M}_{\odot}$. These parameters, however, correspond to $\Gamma \sim 10^{-13}/\text{yr}$, an exceedingly rare event. 
This implies the need for  numerical simulations that account for the non-trivial changes in the cluster structure due to the interaction with the solar system.}

\JCAP{The tidal interaction caused by the Earth is found to leave the cluster relatively intact. This is caused by the quadratic dependence on the disturber mass. Nevertheless, we expect that the crossing of the Earth in the cluster is expected to induce changes in the local structure, for example some turbulence around the Earth location. This would affect the signal and this effect should be studied via some numerical simulations as well.}

Finally, from Eq.~\eqref{ec:kappa} we can also see that a larger axion mass allows for a higher rate because the same value of $\kappa$ can be achieved for a smaller mass of the AMC. Keeping the other parameters fixed, this gives a linear increase in the rate due to the higher number density of the AMCs. However, the achievable increase is somewhat limited due to the axion dark matter mass range as well as the range where spectrally resolved detection is straightforward.

\section{Conclusion}\label{sec:conclusions}

In this paper we have presented a case study in which, by means of a single axion direct detection experiment, it is possible to get insights on the axion-photon coupling and density separately.

Axion haloscopes are, in first instance, only sensitive to the product of coupling and density, $g_{a\gamma\gamma}^2\rho$. Yet, we argue that in the case of an encounter with an axion minicluster (AMC), the energy/frequency spectrum of the power provides additional information on its gravitational potential.
Taking advantage of the good measurement precision of the energy spectrum in haloscope experiments, we are then able to trace the  gravitational potential of an AMC as the Earth goes trough it.
We then use the Poisson equation to connect the extracted gravitational potential to the density of the cluster. 
Combining the information on the density with the power extracted from the haloscope cavity $P\sim g_{a\gamma\gamma}^2\rho$, the axion-photon coupling can be disentangled.

To demonstrate the effectiveness of our method we have applied it to simulated haloscope signals encountering an AMC. This has been done assuming a self-consistent wave function for the miniclusters \cite{Dandoy:2022prp,Yavetz:2021pbc}, as well as incorporating the axion field statistics \cite{Foster:2017hbq,Foster:2020fln}. 

From our simulations, we have extracted the precision of the axion-photon coupling reconstruction based on the number of data points and the AMC parameters. We find that denser miniclusters allow for a better coupling reconstruction, due to their larger gravitational potential and therefore better relative spectral power resolution in the haloscope. We also find that the relative statistical fluctuations of the power are attenuated for denser AMC.

Of course, we have to ask how likely it is to encounter a suitable AMC for which we can indeed reconstruct the axion-photon coupling. Unfortunately, the average rate to cross such an AMC is rather low. If the AMC and axion parameters are of a favorable size the rate can be of the order of one 
\JCAP{per $10^2-10^{3}$~years}\footnote{Larger masses of the axions, or more generally axion-like particles, may allow for an increase in the rate beyond this point, because the reconstruction may be possible for lower minicluster masses that can be more abundant, cf. Eq.~\eqref{ec:kappa}. That said, achieving a sufficient spectral resolution might be difficult at the correspondingly higher frequencies.}. However, it can be much lower. The scaling of which can be inferred from Fig.~\ref{fig:rates_rm} and Eq.~\eqref{ec:kappa}.
\JCAP{Finally, we stress that for our estimates we used a rather simplistic model of a minicluster encountering the detector in a straight line and without being perturbed by the gravitational fields of the Earth and the Sun. In regions with sizable rates this should be a large effect~\cite{Kavanagh:2020gcy,Dandoy:2022prp}. A more careful analysis of the rate should be done, including a detailed simulation of the changes of the minicluster due to its interaction with the solar system and the effects this has on the signal. In light of this our investigation should be taken as an indicative pilot study.}

Let us finally return to the question if we can tell whether axions are a dominant contribution to dark matter. As discussed, after an encounter with a minicluster, we know the coupling. Then, from the measurement of the homogeneous (non-minicluster) dark matter signal\footnote{Estimating the fraction contained in miniclusters is more difficult, as we do not expect to encounter more than one during a reasonable amount of time. Therefore, we cannot obtain a statistically significant result.}, we can measure this part of the density. If this measurement yields a value that is of similar size as the expected density at our location it is at least suggestive that axions are a major component of dark matter. 

\section*{Acknowledgements}

\JCAP{We thank the referee for their very useful and important comments.} We are happy to acknowledge that this project has received support from the European Union’s Horizon 2020 research and innovation programme under the Marie Sklodowska-Curie grant agreement No 860881-HIDDeN.

\appendix

\section{General Formalism of Haloscope Experiment}\label{App: Power in Haloscope }
In the presence of an axion field the Maxwell equations are modified as follows \cite{Sikivie:1983ip,Wilczek:1987mv} \footnote{We assume that no charge or electromagnetic current are present in the resonant cavity.}, 
\begin{equation}\label{Eq: Modified Maxwell equation}
\begin{split}
    &\nabla.\boldsymbol{E}=0 , \\
    &\nabla.\boldsymbol{B}=0 , \\
    &\nabla\times\boldsymbol{E}=-\frac{\partial \boldsymbol{B}}{\partial t} , \\
    &\nabla\times\boldsymbol{B}=\frac{\partial \boldsymbol{E}}{\partial t} - g_{a\gamma\gamma} \boldsymbol{B}\frac{\partial a}{\partial t},
\end{split}    
\end{equation}
where $g_{a\gamma\gamma}$ is the axion photon coupling.\\
In the presence of an external magnetic field $\boldsymbol{B}_0$ permeating the cavity, the axion-photon coupling in the last equations induces an electric field $\boldsymbol{E}_{\rm ind}$. The latter is governed by the following equation of motion, 
\begin{equation}
   (\partial_{t}^{2} - \nabla^{2})\boldsymbol{E_{\text{ind}}}(\boldsymbol{x},t)=g_{a\gamma\gamma}\boldsymbol{B_0}(\boldsymbol{x}) \partial_{t}^{2} a(\boldsymbol{x},t).
   \label{Eq: Electro eq}
\end{equation}
Using the usual expansion of the electric field into cavity modes (cf., e.g.,~\cite{Sikivie:1983ip,Sikivie:1985yu,Melcon:2018dba,Jackson:1998nia})
\begin{equation}
\begin{split}
   &\boldsymbol{E_{\text{ind}}}(\boldsymbol{x},t) =\sum_{j} \alpha_{j}(t)\boldsymbol{E}_{j}(\boldsymbol{x}), \quad \quad \int_V d^3x \lvert \boldsymbol{E}_j(\boldsymbol{x})\rvert^2 = C_j
   \label{Eq: Mode expansion}
  \end{split} 
\end{equation}
where $\boldsymbol{E}_{j}(\boldsymbol{x})$ stands for the mode $j$ and $\alpha_j(t)$ the time dependent coefficient.   
Using this expansion and taking into account damping effects in the cavity, Eq.~\eqref{Eq: Electro eq} can be re-expressed as
\begin{equation}\label{Eq: Induced Electic field}
   (\partial_{t}^{2} + \frac{\omega_j}{Q}\partial_t+\omega_{j}^{2})\alpha_{j}(t)=-b_{j} \partial_{t}^{2} a(\boldsymbol{x},t),
\end{equation}
where $Q$ is the cavity quality factor and with $b_{j}$ is given by\footnote{It is assumed that the typical wavelength of the axion field is much larger than the size of the haloscope such that it is constant in the cavity volume.}
\begin{equation}
    b_{j} = \frac{g_{a\gamma\gamma}}{C_j}\int_V \text{d}^3x \, \boldsymbol{E}^*_{j}.\boldsymbol{B_0}.
\end{equation}
The resulting power extracted from the cavity is finally given in terms of the time averaged electric field and the quality factor of the cavity $Q$ \cite{Melcon:2018dba},
\begin{equation}
\begin{split}
    P_j &= \frac{\omega_{j}}{Q} \frac{1}{2}\int_V \text{d}^3x \, \langle \lvert\alpha_j(t)\rvert^2 \rangle\lvert\boldsymbol{E}_j(\boldsymbol{x})\rvert^2 
\end{split}
\end{equation}
where in the first line, $\langle.\rangle$ denotes the time average and $P_j$ is the power of the mode $j$.\\

In a concrete scenario, $N_T$ measurements of the electric field amplitude $\alpha_{n}(t)$ are taken over a total measurement period $T$ \cite{Foster:2017hbq,Foster:2020fln}. The averaged power extracted from the cavity can then be re-written, 
\begin{equation}\label{Eq: Power to Power Spectral Density}
\begin{split}
    P_j&= \frac{\omega_{j}}{2Q} \int_V \text{d}^3x\, \frac{1}{N_T}\sum_{n=0}^{N_T-1} |\alpha_{j}(t_n) E_{j}(\boldsymbol{x})|^2,\\
    &=\frac{\omega_{j}}{2Q}\frac{1}{N_T^2}\sum_{d=0}^{N_T-1} |\alpha_{j}(w_d)|^2 \int_V\text{d}^3x \,| E_{j}(\boldsymbol{x})|^2,\\
    &=\frac{\omega_{j}}{Q}\frac{1}{4\pi}\sum_{d=0}^{N_T-1} \Delta \omega \frac{T}{N_T^2} |\alpha_{j}(w_d)|^2 \int_V\text{d}^3x \, |E_{j}(\boldsymbol{x})|^2.
\end{split}
\end{equation}
In the second line, the Parseval theorem has been used to relate the time average with a sum over the Fourier modes of the electric field amplitude. For a finite measurement time $T$, only \JCAP{discrete} frequencies enter in the sum, and we have $\Delta \omega=2\pi/T$.\\
The argument of the sum is defined as the spectral power and is explictly given by 
\begin{equation}\label{Eq: Power spectral density}
S(\omega_d) \equiv \frac{T}{N_T^2} |\alpha_{j}(w_d)|^2 \int_V\text{d}^3x \, |E_{j}(\boldsymbol{x})|^2.
\end{equation}
For good enough resolution, the sum in the last line of Eq.~\eqref{Eq: Power to Power Spectral Density} can be approximated by an integral and we have 
\begin{equation}
P_j  \approx \frac{\omega_{j}}{Q}\frac{1}{4\pi} \int d\omega S(\omega).
\end{equation}

\section{Axion minicluster Power Spectral Density}\label{App:AMC PSD}

In this appendix we derive the spectral power for the axion field given in Eq.~\eqref{Eq: Eigenfunctions}. 
We start by solving the differential equation \eqref{Eq: Induced Electic field} for the electric field amplitude $\alpha_j(t)$ and then proceed to calculate its \JCAP{discrete} Fourier transform for a measurement period $T$. From Eq.~\eqref{Eq: Power spectral density} and following the same steps as in Refs.~\cite{Foster:2017hbq,Foster:2020fln}  we obtain the spectral power,
\begin{equation}\label{Eq.App: Power spectral I}
\begin{split}
    S(\omega_d) &= \frac{(\Delta t)^2}{T} \lvert\sum_{nlm}C_{nlm}\sum^{N_T-1}_{n=0}e^{i\omega_d n\Delta t}\left(a_{nlm}   \psi_{nlm}(\boldsymbol{x})  e^{-i\omega_{nlm}n\Delta t} +c.c.\right)\rvert^2 ,\\
    &=\frac{1}{T} \lvert\sum_{nlm}C_{nlm}\sum^{N_T-1}_{n=0} \Delta t \, e^{i\omega_d n\Delta t}\left(a_{nlm}   \psi_{nlm}(\boldsymbol{x})  e^{-i\omega_{nlm}n\Delta t} +c.c.\right)\rvert^2,\\
    &\approx  \frac{1}{T} \lvert\sum_{nlm}C_{nlm}\int^{T/2}_{-T/2} dt \, e^{i\omega_d t}\left(a_{nlm}   \psi_{nlm}(\boldsymbol{x})  e^{-i\omega_{nlm} t} +c.c.\right)\rvert^2,
\end{split}
\end{equation}
where $\Delta t = T/N_T$ and we have introduced the coefficients $C_{nlm}$ as 
\begin{equation}
    C_{nlm}= \sqrt{\left(g_{a\gamma\gamma} B_0\right)^2 \mathcal{G}_j\,V }\frac{\omega_{nlm}^2}{\sqrt{2m_a}\left(\omega_{j}^2-\omega_{nlm}^2-i\omega_{j}\omega_{nlm}/Q\right)}.
\end{equation}
In this last equation $\mathcal{G}_j$ is the usual form factor and is of the order $\mathcal{O}(1)$, $V$ is the cavity volume and $\omega_{nlm}$ is the mode energy \footnote{\JCAP{Note that, compared to Sec.~\ref{sec:minicluster}, we are now in the laboratory frame so that the velocities will be shifted due to a boost by the cluster velocity. We will make this explicit later in the calculation}}. Note that the wave functions $\psi_{nlm}(\boldsymbol{x})$ \JCAP{depend} on the location $\boldsymbol{x}$ (in a frame centered at the origin of the cluster) at which we are doing the measurement in the cluster. We assumed in the main text that the cluster is moving slowly enough so that its motion is neglected during the measurement period $T$.

The time integral can be solved easily noting that
\begin{equation}
    \frac{1}{T}\int^{T/2}_{-T/2} dt\, e^{i\omega_d t} e^{-i\omega_{nlm} t} = \sinc\left(\left(\omega_{nlm}-\omega_d\right)\frac{T}{2}\right).
\end{equation}
With this, Eq.~\eqref{Eq.App: Power spectral I} becomes, 
\begin{equation}\label{Eq.App: Power spectral I}
\begin{split}
    S(\omega_d) \approx  T \lvert\sum_{nlm}C_{nlm} a_{nlm}   \psi_{nlm}(\boldsymbol{x}) \, \sinc\left(\left(\omega_{nlm}-\omega_d\right)\frac{T}{2}\right)\rvert^2.
\end{split}
\end{equation}

The average value of the power spectral density is obtained by taking the average over the random phases. It leads to 
\begin{equation}\label{Eq.App: Power spectral I}
\begin{split}
    \Bar{S}(\omega_d) &= T \langle\lvert\sum_{nlm}C_{nlm} a_{nlm}   \psi_{nlm}(\boldsymbol{x}) \, \sinc\left(\left(\omega_{nlm}-\omega_d\right)\frac{T}{2}\right)\rvert^2\rangle,\\
    &=T \sum_{nlm}\lvert C_{nlm} a_{nlm}  \psi_{nlm}(\boldsymbol{x})\rvert^2 \, \sinc^2\left(\left(\omega_{nlm}-\omega_d\right)\frac{T}{2}\right).
\end{split}
\end{equation}
With the definition of the coefficients $a_{nlm}$ given in Eq.~\eqref{Eq:Coefficients}, we get 
\begin{equation}\label{Eq.App: Power spectral Ia}
\begin{split}
    \Bar{S}(\omega_d) = 4\pi m_a T \int dE \,f(E) \lvert C(E) \rvert^2
  \sqrt{(2m_a\left(E-m_a\phi(r))\right)}\,  \sinc^2\left(\left(E+m_a+\omega_{\rm amc}-\omega_d\right)\frac{T}{2}\right),
\end{split}
\end{equation}
where with the use of the density of states $g(E)$ we have transformed the sum into an integral such that in the continuous limit,  
\begin{equation}
    C(E) =\sqrt{\left(g_{a\gamma\gamma} B_0\right)^2 \mathcal{G}_j\,V }\frac{\left(E+m_a+\omega_{\rm amc}\right)^2}{\sqrt{2m_a}\left(\omega_{j}^2-\left(E+m_a+\omega_{\rm amc}\right)^2-i\omega_{j}\left(E+m_a+\omega_{\rm amc}\right)/Q\right)}.
\end{equation}

\JCAP{However, the energy distribution function obtained  in Sec.~\ref{sec:minicluster} is valid in a frame located at the center of the cluster. Since we are here analyzing the signal in the laboratory frame, the minicluster velocity should be carefully subtracted when the energy is defined. Interpreting  Eq.~\eqref{Eq.App: Power spectral Ia} in a particle picture, we have that the velocities would be shifted as $\boldsymbol{v}\rightarrow \boldsymbol{v}+\boldsymbol{v_c}$, where $\boldsymbol{v_c}$ is the cluster velocity. With this, Eq.~\eqref{Eq.App: Power spectral Ia} becomes,}
\JCAP{\begin{equation}\label{Eq.App: Power spectral I}
\begin{split}
    &\Bar{S}(\omega_d) = m_a^3 T \int d^3 \Tilde{\boldsymbol{v}} \,f(\mid \Tilde{\boldsymbol{v}} -\boldsymbol{v_c}\mid) \lvert C(\mid \Tilde{\boldsymbol{v}} -\boldsymbol{v_c}\mid) \rvert^2\,  \sinc^2\left(\left(m_a+m_a\phi(r)+m_a \Tilde{v}^2/2-\omega_d\right)\frac{T}{2}\right),\\
    &= m_a^3 T \int d\Omega \int_{v\in [0,\sqrt{-2\phi(r)}]}\, d\Tilde{v}\, \Tilde{v}^2  \,f(\mid \Tilde{\boldsymbol{v}} -\boldsymbol{v_c}\mid) \lvert C(\mid \Tilde{\boldsymbol{v}} -\boldsymbol{v_c}\mid) \rvert^2\,  \\
&\qquad\qquad\qquad\qquad\qquad\qquad\qquad\qquad\qquad\times\sinc^2\left(\left(m_a+m_a\phi(r)+m_a \Tilde{v}^2/2-\omega_d\right)\frac{T}{2}\right),
\end{split}
\end{equation}
where we define $\Tilde{\boldsymbol{v}}=\boldsymbol{v}+\boldsymbol{v_c}$ and express the distribution $f(E)$ as a function of the velocity rather than the energy, $E = m_a+m_a\phi(r) + m_a v^2/2$, and include an appropriate transformation of the integration measure. Finally, the integral runs only over laboratory velocities than return velocity in the cluster frame smaller than the escape velocities $v_{\rm max} = \sqrt{-2\phi(r)}$. \\
A further simplification can be made if we use the usual assumption that the distribution function $f(v)$ is constant over the width of the sinc. In this limit, 
\begin{equation}\label{Eq.App: Power spectral I}
\begin{split}
    \Bar{S}(\omega_d) &=4 \pi^2 m_a^2 \Tilde{v}_{d} \int d\theta \sin(\theta) \,f(\Tilde{v}_{d}^2+v_c^2-2\Tilde{v}_{d}v_c\cos(\theta)) \lvert C(\Tilde{v}_{d}^2+v_c^2-2\Tilde{v}_{d}v_c\cos(\theta)) \rvert^2\,\\
    &\qquad\qquad\qquad\times \Theta\left(\sqrt{-2\phi(r)}-(\Tilde{v}_{d}^2+v_c^2-2\Tilde{v}_{d}v_c\cos(\theta))\right) \Theta\left(\Tilde{v}_{d}^2+v_c^2-2\Tilde{v}_{d}v_c\cos(\theta)\right),
\end{split}
\end{equation}
where the velocity $\Tilde{v}_{d}$ returns an energy $\omega_d$, $\Tilde{v}_{d} = \sqrt{2/m_a\left(\omega_d-m_a\phi(r)-m_a\right)}$, $\theta$ is the angle between $\Tilde{\boldsymbol{v}}_{d}$ and $\boldsymbol{v_c}$ and the Heaviside functions ensure that no velocity exceeds the escape velocity.}

\JCAP{Although the last integral over the angle can usually only be performed numerically, it is nevertheless possible to extract the key features of the spectrum: from the Heaviside functions it is easy to see that the signal will be centered around the cluster kinetic energy, $\omega_d = m_a v_c^2/2$. Furthermore, the signal will be contained in the frequency range,
\begin{equation}
    \frac{m_a}{2} v_c^2 +m_a - m_a\sqrt{-2\phi(r)}v_c\leq\omega_d \leq \frac{m_a}{2} v_c^2 +m_a + m_a\sqrt{-2\phi(r)}v_c.
\end{equation}}

\bibliographystyle{JHEP_improved}
\bibliography{./Bib}

\end{document}